\documentclass[zpreprint,zbstepj]{zeus_paper}
\usepackage[english]{babel}

\newcommand{\Ac}{{\cal A}}
\newcommand{\Bc}{{\cal B}}
\newcommand{\Cc}{{\cal C}}
\newcommand{\Dc}{ {\cal D} }
\newcommand{\Ec}{{\cal E}}
\def\citeCTD{{\cite{%
nim:a279:290,*npps:b32:181,*nim:a338:254%
}}\xspace}
\def\citeCAL{{\cite{%
nim:a309:77,*nim:a309:101,*nim:a321:356,*nim:a336:23%
}}\xspace}

\chardef\usc=95
\chardef\til=126
\catcode`\@=11 
\DeclareRobustCommand\xdotspace{\futurelet\@let@token\@xdotspace}
\def\@xdotspace{%
  \ifx\@let@token.\else
  \ifx\@let@token\bgroup.\else
  \ifx\@let@token\egroup.\else
  \ifx\@let@token\/.\else
  \ifx\@let@token\ .\else
  \ifx\@let@token~.\else
  \ifx\@let@token!.\else
  \ifx\@let@token,.\else
  \ifx\@let@token:.\else
  \ifx\@let@token;.\else
  \ifx\@let@token?.\else
  \ifx\@let@token/.\else
  \ifx\@let@token'.\else
  \ifx\@let@token).\else
  \ifx\@let@token-.\else
  \ifx\@let@token\@xobeysp.\else
  \ifx\@let@token\space.\else
  \ifx\@let@token\@sptoken.\else
   .\space
   \fi\fi\fi\fi\fi\fi\fi\fi\fi\fi\fi\fi\fi\fi\fi\fi\fi\fi}
\catcode`\@=12 

\newcommand{\stru}[2]{%
   \relax\ifmmode\hbox{\vrule height#1 depth#2 width0pt}%
   \else\vrule height#1 depth#2 width0pt\fi}

\newcommand{\Ronum}[1]{\uppercase\expandafter{\romannumeral#1}}
\newcommand{\ronum}[1]{\expandafter{\romannumeral#1}}
\DeclareRobustCommand{\LaTeXZ}{%
  \LaTeX\kern-.05em4\kern-.1em
  {\raisebox{-0.2ex}{$\scriptstyle\text{ZEUS}$}}\xspace}

\DeclareMathAlphabet{\mathbf}{OT1}{cmr}{bx}{sl}
\newcommand{\eVdist}{\kern-0.06667em}

\newcommand{\Gev}{{\text{Ge}\eVdist\text{V\/}}}

\newcommand{\gev}{{\,\text{Ge}\eVdist\text{V\/}}}

\newcommand{\Tesla}{\,\text{T}}

\newcommand{\slashfrac}[2]{%
  \raisebox{0.5ex}{\ensuremath #1}\kern-0.12em/\kern-0.08em
  \raisebox{-.8ex}{\ensuremath #2}}

\newcommand{\sqr}[3]{%
    {\vcenter{\hrule height.#3ex\hbox{\vrule width.#2ex height#1ex
     \kern#1ex\vrule width.#3ex}\hrule height.#2ex}}}

\catcode`\@=11 
\newcommand{\parenbar}{\mathpalette\p@renb@r}
\def\p@renb@r#1#2{\vbox{%
  \ifx#1\scriptscriptstyle \dimen@.7em\dimen@ii.2em\else
  \ifx#1\scriptstyle \dimen@.8em\dimen@ii.25em\else
  \dimen@1em\dimen@ii.4em\fi\fi \offinterlineskip
  \ialign{\hfill##\hfill\cr
    \vbox{\hrule width\dimen@ii}\cr
    \noalign{\vskip-.3ex}%
    \hbox to\dimen@{$\mathchar300\hfil\mathchar301$}\cr
    \noalign{\vskip-.3ex}%
    $#1#2$\cr}}}
\catcode`\@=12 

\newcommand{\IP}{{\rm I$\kern-0.01667em$P}\xspace}

\mathchardef\qsm=63
\mathchardef\pls=43
\mathchardef\mns=512
\mathchardef\plm=518
\mathchardef\eql=61
\mathchardef\smallleft=300
\mathchardef\smallright=301
\mathchardef\les=316
\mathchardef\gre=318
\mathchardef\leq=532
\mathchardef\grq=533

\catcode`\@=11 
\newcounter{pict@width}
\newcounter{pict@height}
\newlength{\pict@scale}
\newcommand{\psfigadd}[4]{%
\setcounter{pict@width}{1*\ratio{#2+\pict@scale/2}{\pict@scale}}
\setcounter{pict@height}{1*\ratio{#3+\pict@scale/2}{\pict@scale}}
\setlength{\unitlength}{\pict@scale}
\hbox to #2{\hspace{-\fill}\begin{picture}(\thepict@width,\thepict@height)
\put(0,0){\psfig{figure=#1,width=#2,height=#3,clip=}}
\SetScale{0.283466457}
\SetWidth{1.763889}
{#4}
\end{picture}}
}
\newcounter{pict@widthfst}
\newcounter{pict@widthscd}
\newcounter{pict@widthtot}
\newcommand{\psfigaddtwo}[7]{%
\setcounter{pict@widthfst}{1*\ratio{#2+\pict@scale/2}{\pict@scale}}
\setcounter{pict@widthscd}{1*\ratio{#2+#4+\pict@scale/2}{\pict@scale}}
\setcounter{pict@widthtot}{1*\ratio{#2+#4+#6+\pict@scale/2}{\pict@scale}}
\setcounter{pict@height}{1*\ratio{#3+\pict@scale/2}{\pict@scale}}
\setlength{\unitlength}{\pict@scale}
\hbox{\hspace{-\fill}\begin{picture}(\thepict@widthtot,\thepict@height)
\put(0,0){\psfig{figure=#1,width=#2,height=#3,clip=}}
\put(\thepict@widthscd,0){\psfig{figure=#5,width=#6,height=#3,clip=}}
\SetScale{0.283466457}
\SetWidth{1.763889}
{#7}
\end{picture}}
}
\newcommand{\psfigror}[4]{%
\setcounter{pict@width}{1*\ratio{#2+\pict@scale/2}{\pict@scale}}
\setcounter{pict@height}{1*\ratio{#3+\pict@scale/2}{\pict@scale}}
\setlength{\unitlength}{\pict@scale}
\hbox{\begin{picture}(\thepict@width,\thepict@height)
\put(0,\thepict@height){\psfig{figure=#1,width=#3,height=#2,clip=,angle=270}}
\SetScale{0.283466457}
\SetWidth{1.763889}
{#4}
\end{picture}}
}
\newcommand{\psfigrol}[4]{%
\setcounter{pict@width}{1*\ratio{#2+\pict@scale/2}{\pict@scale}}
\setcounter{pict@height}{1*\ratio{#3+\pict@scale/2}{\pict@scale}}
\setlength{\unitlength}{\pict@scale}
\hbox{\begin{picture}(\thepict@width,\thepict@height)
\put(0,0){\psfig{figure=#1,width=#3,height=#2,clip=,angle=90}}
\SetScale{0.283466457}
\SetWidth{1.763889}
{#4}
\end{picture}}
}
\catcode`\@=12 
\newlength\listtextwidth

\catcode`\@=11 
\newlength{\@tabfninsert}
\newlength{\@tabfnwidth}
\newcommand{\tabfootnote}[2]{%
  \setlength{\@tabfninsert}{0.8em}
  \setlength{\@tabfnwidth}{\textwidth}
  \addtolength{\@tabfnwidth}{-\@tabfninsert}
  \addtolength{\@tabfnwidth}{-0.4em}
  \noindent\makebox[\@tabfninsert][r]{\footnotesize$^{#1}$\hfil}\hfill%
  \parbox[t]{\@tabfnwidth}{\footnotesize #2\hfill}}
\catcode`\@=12 
\begin{document}
\prepnum {DESY 06-129}
\title{
Measurement of azimuthal asymmetries\\
in neutral current deep inelastic scattering
\\
at HERA
}

\author{ZEUS Collaboration}
\date{August, 2006}

\abstract{
The distribution of the azimuthal angle of charged and neutral
hadrons relative to the lepton plane has been studied for neutral 
current deep inelastic $ep$ scattering using an integrated 
luminosity of 45~{\rm pb}$^{-1}$ taken with the ZEUS detector
at HERA. The measurements were made in the hadronic centre-of-mass 
system. The analysis exploits the energy-flow method, which allows 
the measurement to be made over a larger range of pseudorapidity compared 
to previous results. The dependence of the moments of the azimuthal
distributions on the pseudorapidity and minimum transverse energy
of the final-state hadrons are presented. Although the predictions 
from next-to-leading-order QCD describe the data better than do the Monte 
Carlo models incorporating leading-logarithm parton showers, they 
still fail to describe the magnitude of the asymmetries. This 
suggests that higher-order calculations may be necessary to describe 
these data.
}

\makezeustitle
\pagenumbering{Roman}                                                                              
                                                   %
\begin{center}                                                                                     
{                      \Large  The ZEUS Collaboration              }                               
\end{center}                                                                                       
  S.~Chekanov,                                                                                     
  M.~Derrick,                                                                                      
  S.~Magill,                                                                                       
  S.~Miglioranzi$^{   1}$,                                                                         
  B.~Musgrave,                                                                                     
  D.~Nicholass$^{   1}$,                                                                           
  \mbox{J.~Repond},                                                                                
  R.~Yoshida\\                                                                                     
 {\it Argonne National Laboratory, Argonne, Illinois 60439-4815}, USA~$^{n}$                       
\par \filbreak                                                                                     
  M.C.K.~Mattingly \\                                                                              
 {\it Andrews University, Berrien Springs, Michigan 49104-0380}, USA                               
\par \filbreak                                                                                     
  N.~Pavel~$^{\dagger}$, A.G.~Yag\"ues Molina \\                                                   
  {\it Institut f\"ur Physik der Humboldt-Universit\"at zu Berlin,                                 
           Berlin, Germany}
% ...
\newpage                                                                                           
$^{\    1}$ also affiliated with University College London, UK \\                                  
% ...
$^{\dagger}$ deceased \\                                                                           
%
\newpage   
                                                           %
                                                           %
\begin{tabular}[h]{rp{14cm}}                                                                       
$^{a}$ &  supported by the Natural Sciences and Engineering Research Council of Canada (NSERC) \\  
$^{b}$ &  supported by the German Federal Ministry for Education and Research (BMBF), under        
          contract numbers HZ1GUA 2, HZ1GUB 0, HZ1PDA 5, HZ1VFA 5\\                                
$^{c}$ &  supported in part by the MINERVA Gesellschaft f\"ur Forschung GmbH, the Israel Science   
          Foundation (grant no. 293/02-11.2) and the U.S.-Israel Binational Science Foundation \\  
$^{d}$ &  supported by the German-Israeli Foundation and the Israel Science Foundation\\           
$^{e}$ &  supported by the Italian National Institute for Nuclear Physics (INFN) \\                
$^{f}$ &  supported by the Japanese Ministry of Education, Culture, Sports, Science and Technology 
          (MEXT) and its grants for Scientific Research\\                                          
$^{g}$ &  supported by the Korean Ministry of Education and Korea Science and Engineering          
          Foundation\\                                                                             
$^{h}$ &  supported by the Netherlands Foundation for Research on Matter (FOM)\\                   
$^{i}$ &  supported by the Polish State Committee for Scientific Research, grant no.               
          620/E-77/SPB/DESY/P-03/DZ 117/2003-2005 and grant no. 1P03B07427/2004-2006\\             
$^{j}$ &  partially supported by the German Federal Ministry for Education and Research (BMBF)\\   
$^{k}$ &  supported by RF Presidential grant N 1685.2003.2 for the leading scientific schools and  
          by the Russian Ministry of Education and Science through its grant for Scientific        
          Research on High Energy Physics\\                                                        
$^{l}$ &  supported by the Spanish Ministry of Education and Science through funds provided by     
          CICYT\\                                                                                  
$^{m}$ &  supported by the Particle Physics and Astronomy Research Council, UK\\                   
$^{n}$ &  supported by the US Department of Energy\\                                               
$^{o}$ &  supported by the US National Science Foundation\\                                        
$^{p}$ &  supported by the Polish Ministry of Scientific Research and Information Technology,      
          grant no. 112/E-356/SPUB/DESY/P-03/DZ 116/2003-2005 and 1 P03B 065 27\\                  
$^{q}$ &  supported by FNRS and its associated funds (IISN and FRIA) and by an Inter-University    
          Attraction Poles Programme subsidised by the Belgian Federal Science Policy Office\\     
$^{r}$ &  supported by the Malaysian Ministry of Science, Technology and                           
Innovation/Akademi Sains Malaysia grant SAGA 66-02-03-0048\\                                       
\end{tabular}                                                                                      
                                                           %
                                                           %
\pagenumbering{arabic} 
\pagestyle{plain}

\section{Introduction}
\label{sec-int}

\begin{figure}[p]
\vfill
\begin{center}
\epsfig{file=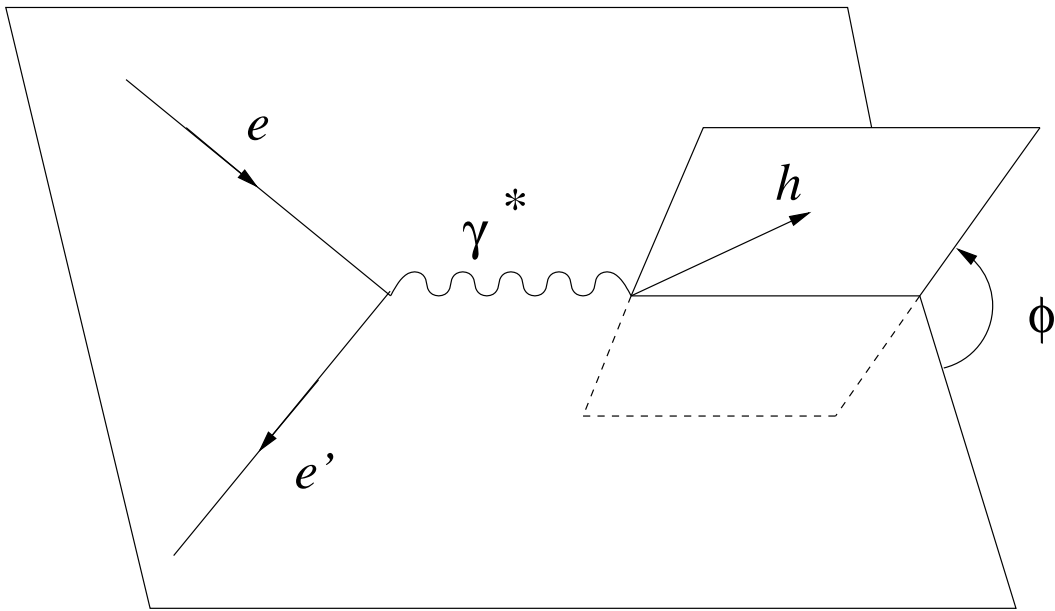,width=14cm,clip=}
\end{center}
\vspace*{1cm}
\caption{The definition of the azimuthal angle $\phi$ either
         in the HCM or the Breit frame. The incoming
         electron is denoted by $e$, the scattered electron by $e'$,
         the exchanged virtual photon by $\gamma^*$ and the
         outgoing hadron or parton by $h$.}
\label{fig-defphi}
\vfill
\end{figure}

The description of the hadronic final state in deep inelastic scattering (DIS) 
is influenced by perturbative Quantum Chromodynamics (pQCD) in several ways 
that can be calculated through exact matrix elements or leading-logarithm 
parton showers. Measurements of the azimuthal distribution of hadrons in 
the semi-inclusive process $e+p\rightarrow e+h+X$ in DIS are sensitive to   
predictions of pQCD. The azimuthal angle, $\phi$, is defined in the 
hadronic centre-of-mass (HCM) frame as the angle between the hadron-production 
plane and the lepton-scattering plane as illustrated in Fig.~\ref{fig-defphi}.

\begin{figure}[p]
\vfill
\unitlength 1mm
 \begin{picture}(150,1)(0,0)
 \put(137,-10){\Large\bf (a)}
 \end{picture}
\psfig{file=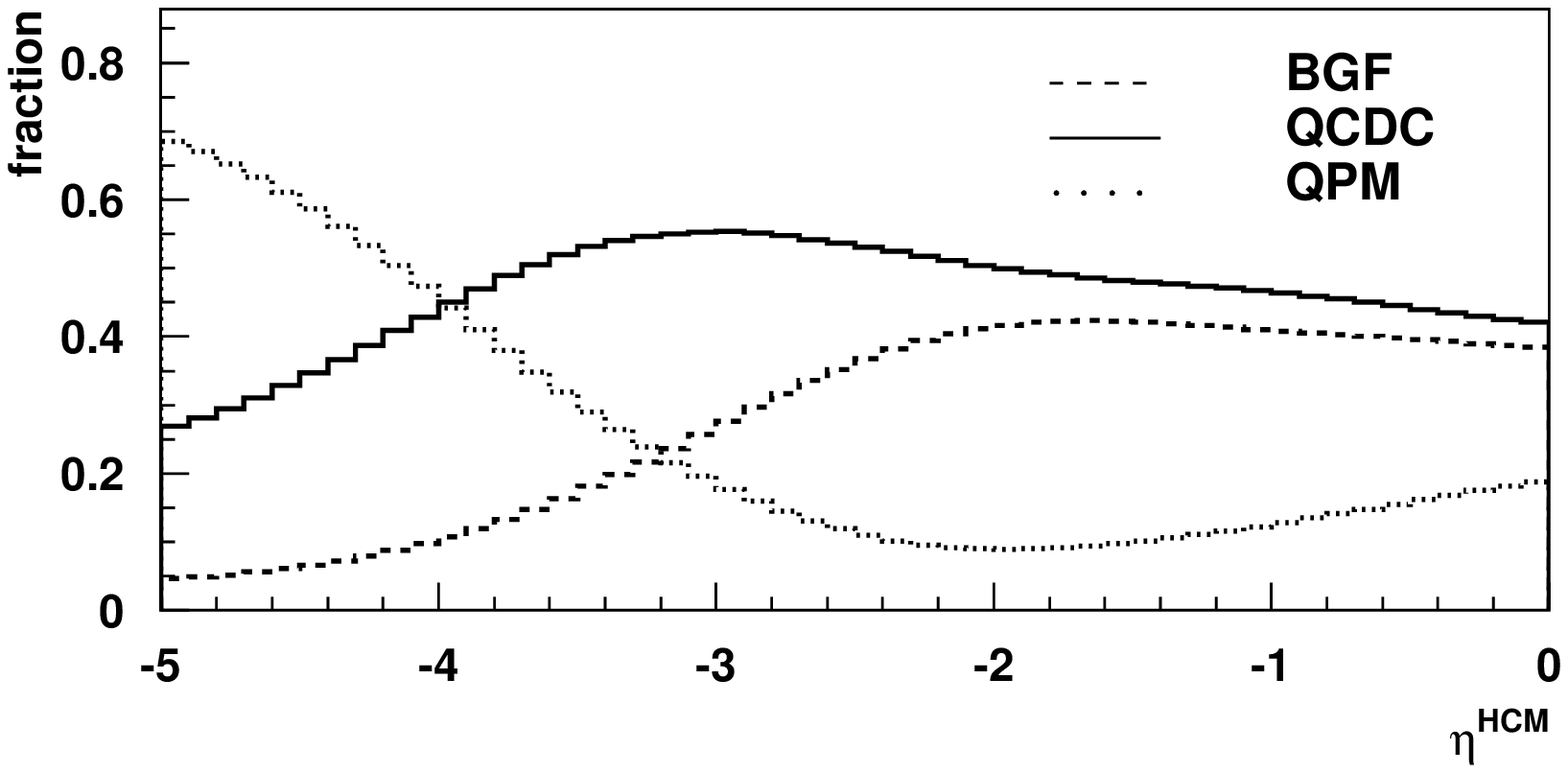,bbllx=19pt,bblly=415pt,bburx=530pt,bbury=658pt,width=15cm,clip=}

\vspace*{1cm}
\unitlength 1mm
 \begin{picture}(150,1)(0,0)
 \put(137,-10){\Large\bf (b)}
 \end{picture}

\hspace*{1.3cm}
\epsfig{file=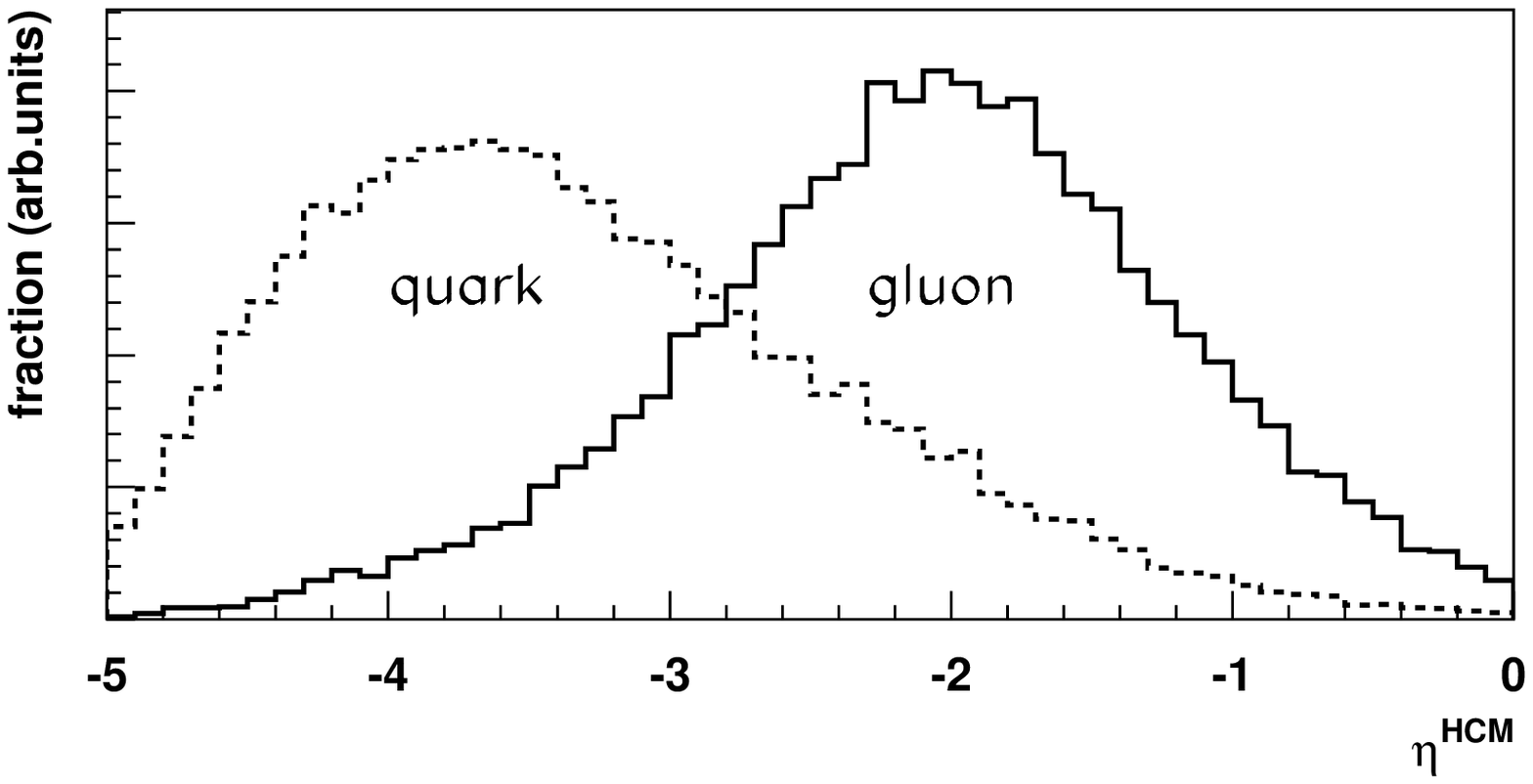,bbllx=50pt,bblly=275pt,bburx=520pt,bbury=520pt,width=13.6cm,clip=}
\vspace*{1cm}
\caption{
(a) The fraction of BGF (dashed line), QCDC (full line) and QPM (dotted line) 
processes as a function of pseudorapidity, $\eta^{HCM}$, in the HCM frame 
for the energy-flow method. (b) For the QCD Compton process, the quark and 
gluon contributions as a function of $\eta^{HCM}$. These predictions were taken 
from {\sc Lepto} 6.5.1 and are shown for the kinematic region 
\mbox{100 $< Q^2 <$ 8000\,GeV$^2$}, \mbox{0.2 $< y <$ 0.8} and $0.01 < x < 0.1$ 
for hadrons with $p_T>0.15\,GeV$ and $\theta > 8^\circ$.
}
\label{fig-qcdc-bgf}
\vfill
\end{figure}

Asymmetries in $\phi$ result from the final-state hadron having transverse 
momentum with respect to the colliding virtual 
photon and the incoming proton. In pQCD, higher-order QCD processes such as QCD 
Compton (QCDC) $(\gamma^* q \rightarrow q g)$ and boson-gluon fusion (BGF)
$(\gamma^* g\rightarrow q \bar{q})$ are the main sources of these hadrons.
These two processes have different $\phi$ behaviours~\cite{GI_HERA91} as well 
as a different pseudorapidity, $\eta$, dependence, defined here with respect to 
the incoming proton direction in the HCM frame. Figure~\ref{fig-qcdc-bgf}(a) 
shows that hadrons from BGF and QCDC dominate over quark-parton-model (QPM) 
$(\gamma^* q \rightarrow q)$ events in the 
region $-4<\eta^{\rm HCM}<0$. In addition, gluons and quarks from the QCDC 
process have different pseudorapidity dependencies, as illustrated in 
Fig.~\ref{fig-qcdc-bgf}(b). 

The azimuthal dependence for semi-inclusive neutral current (NC) DIS can be 
written~\cite{Mendez,pr:d27:84,chay91,ahmed} as:

\begin{equation}
\frac{d\sigma^{ep\rightarrow ehX}}{d\phi}
  = \Ac+\Bc\cos\phi+\Cc\cos2\phi+\Dc\sin\phi+\Ec\sin2\phi.
\label{eq:Xsec}
\end{equation}

The azimuthal asymmetries, specified by the parameters $\Bc,~\Cc,~\Dc$ 
and~$\Ec$, are extracted from the data by calculating the statistical moments 
of the experimental distributions:

\begin{equation}
\langle\cos{\phi}\rangle=\frac{\Bc}{2\Ac};
\hspace*{0.9cm}~~~~~~~~\langle\sin{\phi}\rangle=\frac{\Dc}{2\Ac};
\nonumber
\end{equation}
\begin{equation}
\langle \cos{2\phi}\rangle=\frac{\Cc}{2\Ac};
\hspace*{0.9cm}~~~~~~~~\langle\sin{2\phi}\rangle=\frac{\Ec}{2\Ac}.
\nonumber
\end{equation}

Equation~(\ref{eq:Xsec}) results from the polarisation of the exchanged virtual 
photon. The coefficient $\Bc$ originates from the interference between the 
transversely and longitudinally polarised components; the coefficient $\Cc$ is 
due to the interference of amplitudes corresponding to the $+1$ and $-1$ helicity 
parts of the transversely polarised exchanged boson. The coefficients 
$\Dc$ and $\Ec$ arise from parity-violating weak interactions or longitudinal 
polarisation of the initial lepton beam~\cite{pr:d27:84}.
 They  vanish for purely electromagnetic interactions 
with unpolarised beams. 

It has been proposed~\cite{chay91} to analyse the asymmetry as a 
function of the transverse momentum cutoff, $p_{T}^{\rm cut}$, of a detected 
hadron. Such a cut is efficient in removing QPM events. Consequently, at 
higher $p_{T}^{\rm cut}$ values a better agreement should be obtained with 
the perturbative QCD predictions. A model using a resummation 
formalism~\cite{pl:b515:175} to predict azimuthal asymmetries has also been 
proposed. This model predicts that logarithmic corrections due to soft parton 
emission could be large. A recent paper~\cite{pl:b632:277} showed that a part 
of the asymmetries previously measured by the ZEUS 
collaboration~\cite{pl:b481:199,Asy-jet} may come from terms that are not 
included in the perturbative gluon radiation but are related to the intrinsic 
transverse motion of quarks. Predictions of these models are addressed in this 
paper.

For NC DIS with an unpolarised lepton beam, the 
$\langle\cos\phi\rangle$ and $\langle\cos2\phi\rangle$ values have been measured 
by the ZEUS collaboration~\cite{pl:b481:199,Asy-jet} to be at the few percent 
level. The first publication~\cite{pl:b481:199} measured the azimuthal 
distribution for charged hadrons, whereas the second~\cite{Asy-jet} was performed 
for jets of high transverse energy. The present analysis used the energy-flow 
method, which permits both neutral and charged hadrons to be included in the 
measurements. This analysis was performed using a similar data sample but in an 
extended kinematic range compared to previous publications. In particular, the 
polar-angle range of the measurements 
was increased with respect to the previous studies~\cite{pl:b481:199,Asy-jet}. 
The energy-flow method enhances the contribution of leading hadrons since the 
direction of each particle in the final state is weighted with its transverse 
energy~\cite{zfp:c59:231,*pl:b338:483,*pr:d65:052001,zfp:c63:377,*zfp:c70:609,*epj:c12:595}. Additionally, the values $\langle\sin\phi\rangle$ and 
$\langle\sin 2\phi\rangle$ were determined, although they are 
expected~\cite{pr:d27:84,ahmed} to be much smaller than $\langle\cos\phi\rangle$ 
and $\langle\cos2\phi\rangle$.

This paper presents measurements of $\langle\cos{\phi}\rangle$, 
$\langle \cos{2\phi}\rangle$, $\langle\sin{\phi}\rangle$ and 
$\langle\sin{2\phi}\rangle$ as a function of the pseudorapidity, 
$\eta^{\rm HCM}$, and minimum transverse energy, $E_{T, {\rm min}}^{\rm HCM}$ 
(instead of $p_{T}^{\rm cut}$~\cite{chay91}), of the final-state hadrons.
The results are compared to theoretical expectations.

\section{Data sample}
\label{sec-data}

The experimental results are based on the data collected in 1995-97 with
the ZEUS detector at HERA. Protons of $820\gev$ collided with 
27.5$\gev$ unpolarised\footnote{The positrons were transversely polarised due 
to the Sokolov-Ternov~\cite{spd:8:1203} effect, but had no longitudinal 
polarisation.} positrons. Neutral current DIS events were 
selected from data corresponding to an integrated luminosity 
of $45~{\rm pb}^{-1}$. A detailed description of the ZEUS detector can be found 
elsewhere~\cite{zeus:1993:bluebook}. A brief outline of the components that are 
most relevant for this analysis is given below.

The high-resolution uranium--scintillator calorimeter (CAL)~\citeCAL consists
of three parts: the forward (FCAL), the barrel (BCAL) and the rear (RCAL)
calorimeters. Each part is subdivided transversely into towers and
longitudinally into one electromagnetic section (EMC) and either one (in RCAL)
or two (in BCAL and FCAL) hadronic sections (HAC). The smallest subdivision of
the calorimeter is called a cell.  The CAL energy resolutions, as measured under
test-beam conditions, are $\sigma(E)/E=0.18/\sqrt{E}$ for electrons and
$\sigma(E)/E=0.35/\sqrt{E}$ for hadrons, with $E$ in $\Gev$.

Charged particles are tracked in the central tracking detector (CTD)~\citeCTD,
which operates in a magnetic field of $1.43\Tesla$ provided by a thin
superconducting coil. The CTD consists of 72~cylindrical drift chamber
layers, organised in 9~superlayers covering the polar-angle\footnote{The
ZEUS coordinate system is 
a right-handed Cartesian system, with the $Z$
axis pointing in the proton beam direction, referred to as the ``forward
direction'', and the $X$ axis pointing left towards the centre of HERA.
The coordinate origin is at the nominal interaction point.\xspace}
 region
\mbox{$15^\circ<\theta<164^\circ$}. The transverse-momentum resolution for
full-length tracks is $\sigma(p_T)/p_T=0.0058p_T\oplus0.0065\oplus0.0014/p_T$,
with $p_T$ in $\Gev$.

The selection criteria were based on an earlier ZEUS 
investigation~\cite{pl:b481:199}. The main requirements on the event were:

\begin{itemize}

\item an identified scattered positron with energy $E'_{e}>10{\rm ~GeV}$. 
      Energy deposits in the CAL, within an $\eta-\phi$ cone of radius 1 and 
      consistent with being a photon, were removed from the calculation of 
      the positron energy;

\item $100<Q^2<8000{\rm ~GeV^2}$, $0.2<y<0.8$ and $0.01<x<0.1$. The quantities 
      $x,y$ and $Q^2$ are respectively $x$-Bjorken, the inelasticity, and the 
      negative square of the exchanged boson virtuality. The double angle 
      method was used to reconstruct these variables and so determine 
      the direction of the exchanged boson~\cite{proc:hera:1991:23,*proc:hera:1991:43}. 
      The mean values of $x$, $y$ and $Q^2$ were respectively $0.0222$, $0.351$ and 
      700\,GeV$^2$. The final sample consisted of 16\,472 events.

\end{itemize}

Charged and neutral final-state particles were reconstructed using a combination 
of track and calorimeter information that optimises the resolution of the 
reconstructed kinematic variables~\cite{epj:c6:43,*briskin:phd:1998}. The 
selected tracks and calorimeter clusters are referred to as energy flow objects 
(EFOs). The EFOs were required to satisfy the following:

\begin{itemize}

\item transverse momentum in the laboratory frame $p_T>$ 0.15\,GeV;

\item polar angle $\theta > 8^\circ$.

\end{itemize}

These cuts ensured the analysis was performed in a region of high acceptance and 
high detector efficiency. An average of about 18 EFOs per event were 
reconstructed, yielding a total of 293\,000 objects, each of which provided a 
value of the azimuthal angle, $\phi$, used for further analysis.

\section{Method}
\label{sec-method}

The energy-flow method was proposed~\cite{np:b162:125,np:b206:239} to ensure that 
the inclusive variables such as those measured here are infra-red and collinear 
safe due to the direction of each particle being weighted by its transverse energy.

In this analysis, the moments of trigonometric functions of the azimuthal angle are 
calculated as follows. For a function, $F(n \phi^{\rm HCM})$, where 
$F(n \phi^{\rm HCM})$ can be $\sin(n \phi^{\rm HCM})$ or $\cos(n \phi^{\rm HCM})$ and 
$n = 1$ or 2, the mean value was determined using the formula:

\begin{equation}
\langle F\left(n \phi^{\rm HCM}\right) \rangle = \frac{\sum_i E_{T, i}^{\rm HCM} 
                                            F\left(n \phi_i^{\rm HCM}\right)}
					   {\sum_i E_{T, i}^{\rm HCM}}, 
\label{eq:method}
\end{equation}
 
where $E_T^{\rm HCM}$ is the transverse energy and $\phi$ the azimuthal angle for  
each EFO, $i$. The value is calculated in bins of 
$\eta^{\rm HCM}$, semi-inclusively, and for different minimum cuts on the transverse 
energy of the EFO, $E_{T, {\rm min}}^{\rm HCM}$. Equation~(\ref{eq:method}) is also 
used for the determination of the moments in theoretical calculations, in which the 
sum is either over final-state hadrons or partons. The  mean values are not expected 
to be sensitive to uncertainties in parton distribution functions, fragmentation 
functions, and calorimeter energy scale, since such effects contribute to both the 
numerator and the denominator.

\section{Correction procedure}
\label{sec-corrections}

Monte Carlo (MC) events were used to correct the data for detector acceptances. 
For all generated events, the ZEUS detector response was simulated in detail using 
a program based on {\sc Geant} 3.13~\cite{tech:cern-dd-ee-84-1}.

Neutral current events with electroweak radiative corrections were simulated 
with the {\sc Lepto} 6.5.1~\cite{cpc:101:108} program interfaced to 
{\sc Heracles} 4.6.1~\cite{cpc:69:155,spi:www:heracles} via the {\sc Djangoh} 1.1
program~\cite{spi:www:djangoh11,cpc:81:381}. High-order QCD processes were 
simulated using the MEPS option of {\sc Lepto}. A second sample of MC 
events was generated with {\sc Ariadne} 4.12~\cite{cpc:71:15}, where 
the QCD cascade is simulated with the colour-dipole model. In all cases, the 
events were generated using the CTEQ4D parton density parametrisation\cite{lai} 
of the proton. The final-state partonic system was hadronised using the Lund string 
model as implemented in {\sc Jetset} 7.4~\cite{cpc:43:367}.

For a given bin, $j$, in $\eta^{\rm HCM}$ and $F(n \phi^{\rm HCM})$, two correction 
factors for $F(n \phi^{\rm HCM})$ were derived from MC as the ratio of values 
determined using hadrons and EFOs:

$$
C_j = \frac{\left[\sum_i E_{T, i}^{\rm HCM} F\left(n \phi_i^{\rm HCM}\right)\right]^{\rm MC}_{{\rm had}, j}}
         {\left[\sum_i E_{T, i}^{\rm HCM} F\left(n \phi_i^{\rm HCM}\right)\right]^{\rm MC}_{{\rm EFO}, j}}, 
\ \ \ \ \ D_j = \frac{\left[\sum_i E_{T, i}^{\rm HCM}\right]^{\rm MC}_{{\rm had}, j}}
         {\left[\sum_i E_{T, i}^{\rm HCM}\right]^{\rm MC}_{{\rm EFO}, j}},
$$

where $C_j$ corrects the trigonometric function weighted with $E_T$ and $D_j$ corrects 
the sum of $E_T$ in the bin, $j$. The corrected data for a bin in $\eta^{\rm HCM}$ is 
then given by:

$$
\langle F\left(n \phi^{\rm HCM}\right) \rangle 
= \frac{\sum_{j} \left(\sum_i E_{T, i}^{\rm HCM} F\left(n \phi_i^{\rm HCM}\right) \right)_{{\rm EFO}, j} \cdot C_j}
       {\sum_{j} \left(\sum_i E_{T, i}^{\rm HCM} \right)_{{\rm EFO}, j} \cdot D_j }.
$$ 

The overall correction factors are about 10\% and arise mainly from undetected hadrons.

For this approach to be valid, the uncorrected energy flow in the data must be well 
described by the MC simulations at the detector level. This condition was 
satisfied~\cite{thesis:ukleja:2006} 
by both the {\sc Lepto} and {\sc Ariadne} simulations in the $\rm \eta^{HCM}$ and 
$E_T^{\rm HCM}$ regions under investigation. The samples of {\sc Lepto} events were 
used for the final corrections.

Systematic uncertainties of the azimuthal asymmetry were determined by varying the 
event selection cuts within their reconstruction resolution and the total systematic 
uncertainty was taken as the sum in quadrature of the individual systematics. The 
dominant contributions originated from the following sources (maximal deviations for 
$\langle \cos\phi \rangle$ are shown in parentheses):

\begin{itemize}

\item the use of the {\sc Ariadne} MC model to correct the data (0.017);

\item varying the cut on the $p_T$ of the final-state objects from 0.15 GeV to 0.2 GeV 
      to estimate the effect due to low-$p_T$ tracks (0.013);

\item the inclusion of energy deposits consistent with being a photon for the 
      calculation of the hadronic angle, used in the double angle 
      method~\cite{proc:hera:1991:23,*proc:hera:1991:43}, and therefore the 
      transformation to the HCM frame (0.009).

\end{itemize}

The effect of the variation of other selection cuts was negligible.

\section{QCD calculations}
\label{sec-mc}

The data were compared to the MC programs {\sc Lepto} and {\sc Ariadne}, described 
in the previous section, and to a next-to-leading-order (NLO) prediction. In these two 
MC programs, the azimuthal-angular distribution is implemented according to the 
first-order QCD matrix elements for QCDC and BGF. The parton-shower and 
soft-matrix-element events are flat in azimuth. The NLO 
predictions were calculated using the dipole factorisation formulae~\cite{disent1} 
implemented in the {\sc Disent} program~\cite{disent1,disent2}. The calculations 
used a generalised version of the subtraction method~\cite{nlo-sub} and were 
performed in the massless $\overline{\rm MS}$ renormalisation and factorisation 
schemes. The azimuthal-angle distribution is calculated to NLO by the {\sc Disent} 
program through the use of the photon leptonic current in the amplitudes for the QCDC, 
BGF and $\mathcal{O}(\alpha_s^2)$ processes. This program contains neither $Z^0$ 
exchange nor hadronisation effects. The following 
settings were used as defaults for {\sc Disent}: the number of flavours was  
five, the factorisation and renormalisation scales were $\mu_{F,R}=Q$, 
and the parton distribution function was CTEQ3M\cite{pl:b304:159}. Samples 
of events from {\sc Lepto} 6.5.1 
were used to correct the NLO QCD calculations for $Z^0$-exchange effects, hadronisation 
and undetected particles due to the requirements on $p_T$ and $\theta$ of the produced 
hadrons. 

The uncertainty in the {\sc Disent} predictions was estimated by changing the following:

\begin{itemize}

\item the renormalisation and factorisation scales were individually changed to 
      $\mu_{F,R}=Q/2$ and $2Q$;

\item the parton distribution functions CTEQ4M~\cite{lai}, CTEQ5M~\cite{epj:c12:375}, 
      were used;

\item the correction for $Z^0$-exchange effects, hadronisation and undetected particles 
      was repeated with the {\sc Ariadne} 4.12 MC.

\end{itemize}

The above uncertainties were at most 0.004 in both $\langle \cos\phi \rangle$ 
and $\langle \cos2\phi \rangle$. They were added in quadrature and are displayed in the 
figures as shaded bands around the central prediction. The sensitivity to different gluon 
distributions was checked by using the MRST99~\cite{epj:c4:463,*epj:c14:133} parton 
distributions functions with an increased or decreased gluon density; the differences 
were negligible.

\section{Results}
\label{sec-res}

\begin{figure}[p]
\unitlength 1mm
\setlength{\unitlength}{1mm}
\centering
\begin{picture}(80,100)(0,0)
 
\put(-55,0){\epsfig{file=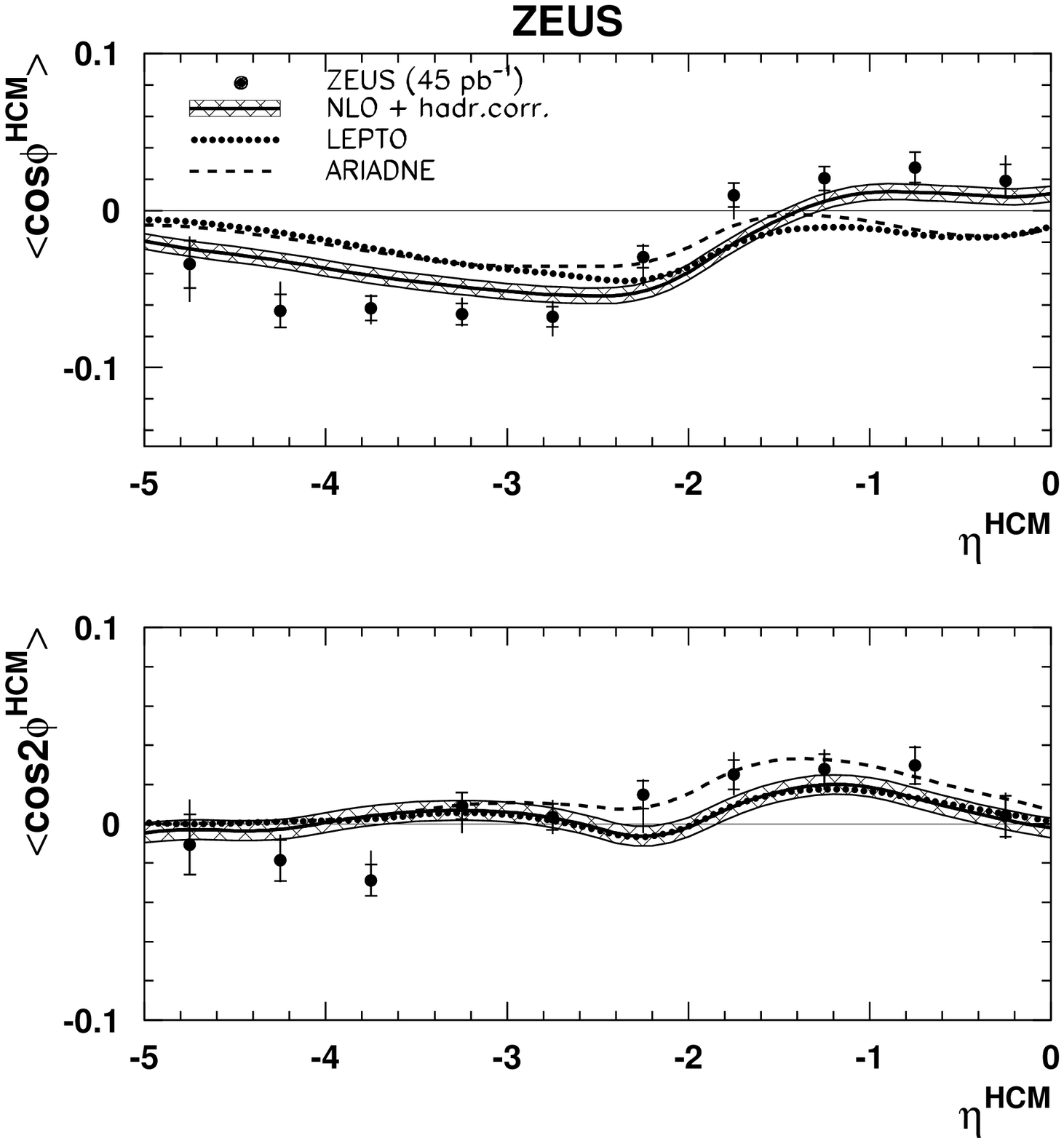,width=9cm,clip=}}
 
\put(40,0){\epsfig{file=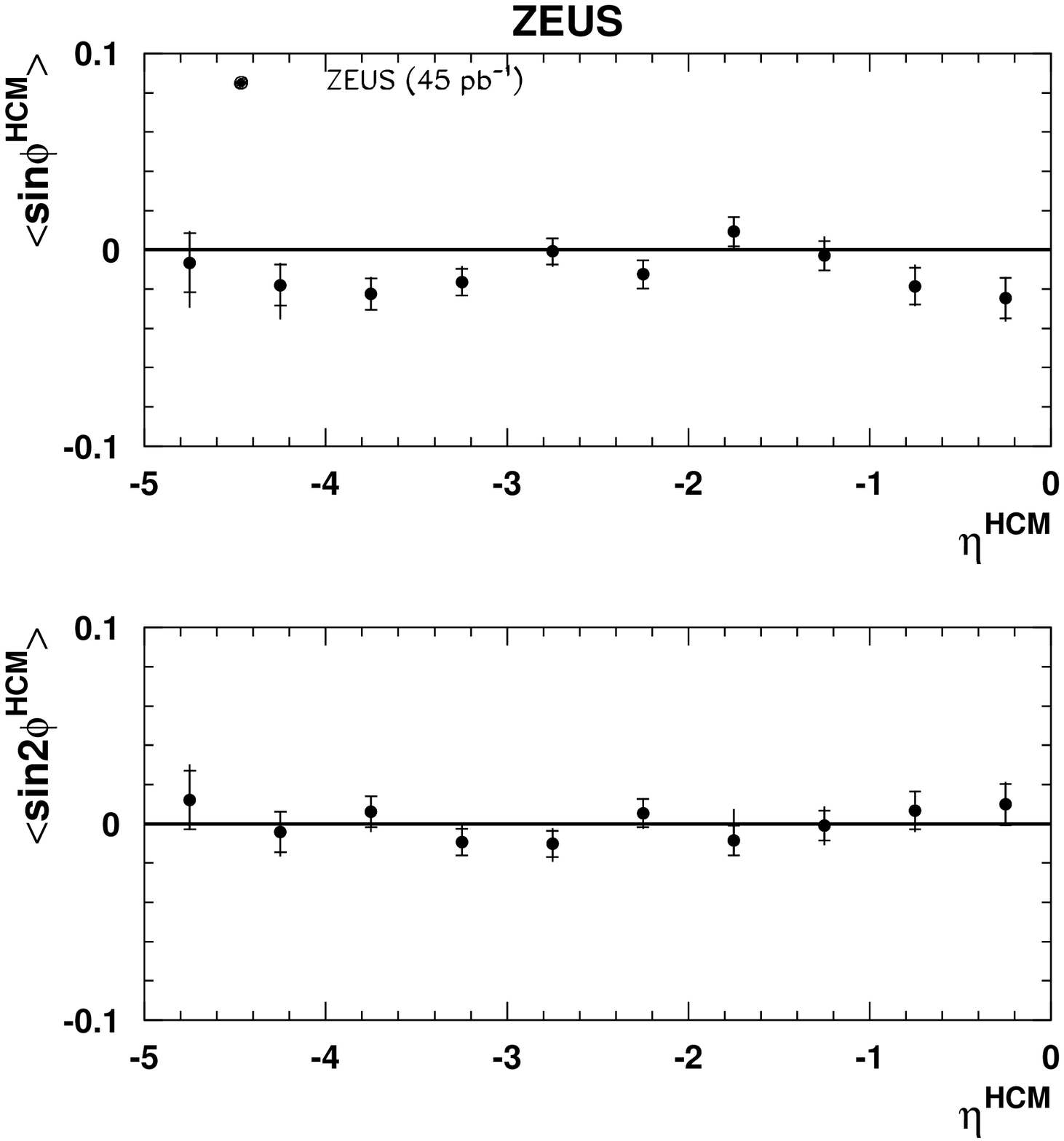,width=9cm,clip=}}
\end{picture}
\caption{The values of $\langle\cos\phi^{HCM}\rangle$, 
         $\langle\cos2\phi^{HCM}\rangle$, $\langle\sin\phi^{HCM}\rangle$
         and $\langle\sin2\phi^{HCM}\rangle$, calculated using the energy-flow method 
	 as in Eq.~(\ref{eq:method}), as a function of hadron pseudorapidity, 
	 $\eta^{HCM}$. They were obtained in the HCM frame 
	 for the kinematic region \mbox{100 $< Q^2 <$ 8000\,GeV$^2$}, $0.01<x<0.1$ 
	 and $0.2<y<0.8$ for hadrons with $p_T>0.15\,GeV$ and $\theta > 8^\circ$. 
	 The inner error bars are statistical uncertainties, the 
	 outer are statistical and systematic uncertainties added in quadrature. The 
	 NLO QCD predictions of {\sc Disent} (solid line), with its associated uncertainty 
	 (shaded band), corrected for hadronisation and hadron losses (see text), the 
	 predictions of {\sc Lepto} 6.5.1 (dotted line), and the predictions of 
	 {\sc Ariadne} 4.12 (dashed line) are shown.}
\label{fig-phihcm_etahcm}
\end{figure}

Azimuthal asymmetries have been measured in NC DIS events with the requirements:   
\mbox{100 $< Q^2 <$ 8000 GeV$^2$}, \mbox{0.2 $< y <$ 0.8} and $0.01 < x < 0.1$ for 
hadrons with $p_T>0.15$\,{\rm GeV} and $\theta > 8^\circ$. The mean 
values of $\rm \cos \phi^{HCM}$, $\rm \cos 2\phi^{HCM}$, $\rm \sin\phi^{HCM}$ 
and  $\rm \sin 2\phi^{HCM}$ are shown in Fig.~\ref{fig-phihcm_etahcm} and 
given in Tables~\ref{tab-asym.cos} and~\ref{tab-asym.sin} as a 
function of $\eta^{\rm HCM}$. The data are compared with predictions 
from MC models and from NLO QCD as described in the previous section.

Figure~\ref{fig-phihcm_etahcm} shows that the value of 
$\langle\cos{\phi^{\rm HCM}}\rangle$ is negative for $\eta^{\rm HCM}<-2$ but 
becomes positive for larger $\eta^{\rm HCM}$. This is in disagreement with both 
MC predictions, which are less negative for $\eta^{\rm HCM}<-2$ and remain 
negative for larger $\eta^{\rm HCM}$. The measured 
$\langle\cos{2\phi^{\rm HCM}}\rangle$ values are consistent with zero for 
$\eta^{\rm HCM}<-2$ but are positive for higher values of $\eta^{\rm HCM}$. 
This is consistent with the expectations from both {\sc Lepto} and {\sc Ariadne}.

The NLO QCD predictions, corrected for hadronisation, agree better with the 
experimental values for $\langle\cos{\phi^{\rm HCM}}\rangle$ than do the MC 
predictions. The predictions from NLO QCD are more negative for 
\mbox{$\eta^{\rm HCM}<-2$} than those from the MC generators and also have 
positive values for larger $\eta^{\rm HCM}$. However, the NLO calculation still 
fails to describe the magnitude of the asymmetry in the data. The comparison 
with NLO QCD was also made at higher $p_T$, greater than 1\,GeV (not shown). 
The Monte Carlo was used to correct the NLO for this cut; although the final 
correction for $\langle\cos{\phi^{\rm HCM}}\rangle$ was small, the correction 
for hadron removal was large. However, the comparison with the data was 
qualitatively the same as when the more inclusive cut was used. The disagreement 
between data and NLO suggests that higher-order calculations may be necessary 
to describe this distribution fully. Inclusion of higher orders through a 
resummation of large logarithmic terms is expected~\cite{pl:b515:175} to give 
an improved description compared to that of LO for $-5 < \eta < -3$. However, the 
description is not significantly better than for the other predictions. For 
$\langle\cos{2\phi^{\rm HCM}}\rangle$, the NLO and MC predictions are similar 
and describe the data reasonably well.

Figure~\ref{fig-phihcm_etahcm} shows that the values of
$\langle\sin{\phi^{\rm HCM}}\rangle$ and $\langle\sin{2\phi^{\rm HCM}}\rangle$
are small. A deviation of $\langle\sin{\phi^{\rm HCM}}\rangle$ from zero at the 
level of three standard deviations is observed. The mean values are expected to 
be at least an order of magnitude smaller than the 
$\langle\cos{\phi^{\rm HCM}}\rangle$ term~\cite{ahmed}. The values of 
$\langle\sin{2\phi^{\rm HCM}}\rangle$ are consistent with zero. None of the 
theoretical models include predictions for 
$\langle\sin{\phi^{\rm HCM}}\rangle$ or $\langle\sin{2\phi^{\rm HCM}}\rangle$.

\begin{figure}[htp]
\begin{center}
\epsfig{file=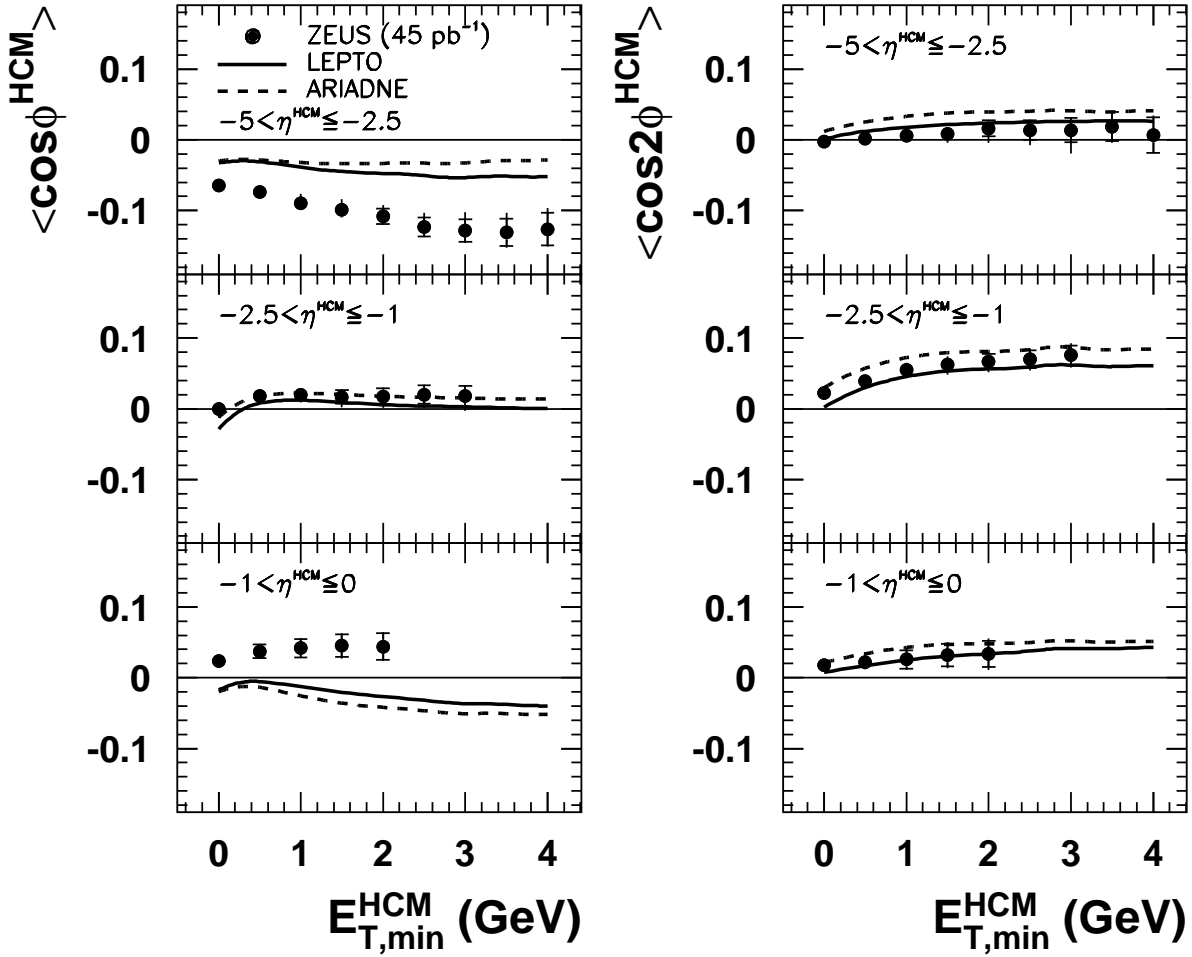,width=16cm,clip=}
\end{center}
\caption{The values of $\langle\cos\phi^{HCM}\rangle$ and 
         $\langle\cos2\phi^{HCM}\rangle$, calculated using the energy-flow method 
         as in Eq.~(\ref{eq:method}), as a function of hadron minimum transverse energy,  
	 $E_{T,min}^{HCM}$. They were obtained in the HCM frame for the pseudorapidity 
	 intervals $-5<\eta^{HCM}\leq -2.5$, $-2.5<\eta^{HCM}\leq -1$ and 
	 $-1<\eta^{HCM}\leq 0$ in the kinematic region \mbox{100 $< Q^2 <$ 8000\,GeV$^2$}, 
	 $0.01<x<0.1$ and $0.2<y<0.8$ for hadrons with $p_T>0.15\,GeV$ and $\theta > 8^\circ$.
         The inner error bars are statistical uncertainties, the outer are statistical 
	 and systematic uncertainties added in quadrature. The predictions of 
	 {\sc Lepto}~6.5.1 (solid line) and of {\sc Ariadne}~4.12 (dashed line) are 
	 shown.
}
\label{fig-cosphihcm_ethcm}
\end{figure}

To investigate the effect of the minimum transverse energy cut, 
$E_{T,{\rm min}}^{\rm HCM}$, on the asymmetries, the event sample was subdivided 
into three regions of $\eta^{\rm HCM}$: $-5<\eta^{\rm HCM}<-2.5 $, 
$-2.5<\eta^{\rm HCM}<-1$ and $-1<\eta^{\rm HCM}<0$. For 
$E_{T,{\rm min}}^{\rm HCM} = 1$\,GeV, the acceptance is approximately 100\%. 
Below this value, some hadrons are removed by the $p_T>$ 0.15\,GeV requirement 
mainly in the region $-2.5<\eta^{\rm HCM}<-1$. The data are shown in 
Fig.~\ref{fig-cosphihcm_ethcm}, and given in Tables~\ref{tab-ethcm.cos} 
and~\ref{tab-ethcm.cos2}, compared to the predictions from {\sc Lepto} and 
{\sc Ariadne} MCs. As stated previously, NLO QCD predictions for higher 
$E_{T,{\rm min}}^{\rm HCM}$ have large corrections for hadron removal.

The first region $-5<\eta^{\rm HCM}<-2.5 $ is  part of the current region in DIS 
defined in the Breit frame as $\eta^{\rm Breit}\approx \eta^{\rm HCM}+2<0$; in 
this region the main contribution to the azimuthal asymmetry comes from QCDC and 
arises from hadrons from quark fragmentation (Fig.~\ref{fig-qcdc-bgf}).
This region was investigated in the first ZEUS analysis of azimuthal 
asymmetries~\cite{pl:b481:199} using charged hadrons. The data from the current 
analysis, shown in Fig.~\ref{fig-cosphihcm_ethcm}, confirm, with higher experimental 
precision and to higher $E_{T,{\rm min}}^{\rm HCM}$, that the value of 
$\langle\cos{\phi^{\rm HCM}}\rangle$ is more negative than expected from MC 
predictions. The $\langle\cos{2\phi^{\rm HCM}}\rangle$ values are small and 
consistent with zero and also in agreement with both {\sc Lepto} and {\sc Ariadne}.

The region $-2.5<\eta^{\rm HCM}<-1$ is still dominated by QCDC events but the 
contribution from BGF events increases. 
The ZEUS analysis of azimuthal asymmetries measured using jets\cite{Asy-jet} was 
based on hadrons from this region of phase space. In that analysis, a large 
positive value of $\langle\cos{2\phi^{\rm HCM}}\rangle$ was measured, whereas the 
$\langle\cos{\phi^{\rm HCM}}\rangle$ value was consistent with zero. The 
$\langle\cos{2\phi^{\rm HCM}}\rangle$ values agreed with the NLO QCD prediction and 
were inconsistent with the LO prediction. The  results presented here in
 Fig.~\ref{fig-cosphihcm_ethcm} confirm, with higher experimental precision, a small value of 
$\langle\cos{\phi^{\rm HCM}}\rangle$ and positive values for 
$\langle\cos{2\phi^{\rm HCM}}\rangle$ for all $E_{T,{\rm min}}^{\rm HCM}$. 
The MC predictions of {\sc Lepto} and {\sc Ariadne} are in good agreement with 
the data.

The third region, $-1<\eta^{\rm HCM}<0$, is populated roughly equally by hadrons 
from QCDC and from BGF processes. This measurement extends the kinematic region 
over those presented previously. The results are presented in 
Fig.~\ref{fig-cosphihcm_ethcm}. The $\langle\cos{\phi^{\rm HCM}}\rangle$ values 
are positive, contrary to MC predictions, whereas the 
$\langle\cos{2\phi^{\rm HCM}}\rangle$ values are positive and in agreement with 
MC predictions. These trends persist up to the highest $E_{T,{\rm min}}^{\rm HCM}$ 
values measured.

\section{Summary and conclusions}
\label{sec-conclusions}

The azimuthal asymmetries in  deep inelastic scattering have been measured in the 
hadronic centre-of-mass frame at HERA for a selected sample of neutral current 
events with  $100<Q^2<8000$~GeV$^2$, $0.2<y<0.8$ and $0.01<x<0.1$. An energy-flow 
analysis method was used, which permitted the use of both neutral and charged hadrons 
and extends the phase space over previous measurements. 

Azimuthal asymmetries have been investigated as a function of hadron pseudorapidity,  
$\eta^{\rm HCM}$. The value of $\langle\cos{\phi^{\rm HCM}}\rangle$ is negative 
for $\eta^{\rm HCM}<-2$ but becomes positive for larger $\eta^{\rm HCM}$. The 
distribution is not well described by the MC predictions of {\sc Ariadne} and 
{\sc Lepto}. Although the predictions from NLO QCD describe the data better than 
do the MCs, they still fail to describe the magnitude of the asymmetries. This 
suggests 
that higher-order calculations may be necessary to describe these data. However, 
the predicted values of $\langle\cos{2\phi^{\rm HCM}}\rangle$ in both MC models 
and in NLO QCD agree with the data. A deviation of 
$\langle\sin{\phi^{\rm HCM}}\rangle$ from zero at the level of three standard 
deviations is observed. The values of $\langle\sin{2\phi^{\rm HCM}}\rangle$ are 
consistent with zero.

\section*{Acknowledgements}
We thank the DESY Directorate for their strong support and encouragement.
The effort of the HERA machine group is gratefully acknowledged.
We thank the DESY computing and network services for their support
as well as the engineers and technicians for their work in constructing
and maintaining the ZEUS detector. Special thanks go to Thomas Gehrmann, 
Gunnar Ingelman, Krzysztof Kurek, Piet Mulders and Pavel Nadolsky for 
valuable discussions. 

\vfill\eject

\providecommand{\etal}{et al.\xspace}
\providecommand{\coll}{Collab.\xspace}
\catcode`\@=11
\def\@bibitem#1{%
\ifmc@bstsupport
  \mc@iftail{#1}%
    {;\newline\ignorespaces}%
    {\ifmc@first\else.\fi\orig@bibitem{#1}}
  \mc@firstfalse
\else
  \mc@iftail{#1}%
    {\ignorespaces}%
    {\orig@bibitem{#1}}%
\fi}%
\catcode`\@=12


\newpage

\renewcommand{\arraystretch}{1.1}

\begin{table}[!htb]
\begin{center}
\begin{tabular}{||c|c c c ||}
\hline
$\eta^{\rm HCM}$ &$\langle\cos{\phi^{\rm HCM}}\rangle$
   & $\delta_{\rm stat}$ & $\delta_{\rm syst}$  \\
\hline \hline
$-4.75$ & $-0.034$ & $\pm 0.015$ & $_{-0.019}^{+0.009}$  \\
$-4.25$ & $-0.064$ & $\pm 0.010$ & $_{-0.003}^{+0.016}$  \\
$-3.75$ & $-0.062$ & $\pm 0.008$ & $_{-0.006}^{+0.004}$  \\
$-3.25$ & $-0.066$ & $\pm 0.007$ & $_{-0.004}^{+0.008}$  \\
$-2.75$ & $-0.068$ & $\pm 0.006$ & $_{-0.011}^{+0.008}$  \\
$-2.25$ & $-0.030$ & $\pm 0.007$ & $_{-0.017}^{+0.005}$  \\
$-1.75$ & $~0.010$ & $\pm 0.008$ & $_{-0.013}^{+0.002}$  \\
$-1.25$ & $~0.020$ & $\pm 0.008$ & $_{-0.012}^{+0.002}$  \\
$-0.75$ & $~0.028$ & $\pm 0.010$ & $_{-0.007}^{+0.002}$  \\
$-0.25$ & $~0.019$ & $\pm 0.010$ & $_{-0.004}^{+0.012}$  \\
\hline
\multicolumn{4}{c}{~~~}\\
\hline
$\eta^{\rm HCM}$ &$\langle\cos{2\phi^{\rm HCM}}\rangle$
   & $\delta_{\rm stat}$ & $\delta_{\rm syst}$  \\
\hline \hline
$-4.75$ & $-0.011$ & $\pm 0.015$ & $_{-0.002}^{+0.017}$  \\
$-4.25$ & $-0.019$ & $\pm 0.010$ & $_{-0.003}^{+0.008}$  \\
$-3.75$ & $-0.029$ & $\pm 0.008$ & $_{-0.002}^{+0.013}$  \\
$-3.25$ & $~0.009$ & $\pm 0.007$ & $_{-0.012}^{+0.000}$  \\
$-2.75$ & $~0.004$ & $\pm 0.007$ & $_{-0.006}^{+0.006}$  \\
$-2.25$ & $~0.015$ & $\pm 0.007$ & $_{-0.018}^{+0.004}$  \\
$-1.75$ & $~0.025$ & $\pm 0.008$ & $_{-0.007}^{+0.009}$  \\
$-1.25$ & $~0.028$ & $\pm 0.008$ & $_{-0.008}^{+0.006}$  \\
$-0.75$ & $~0.030$ & $\pm 0.009$ & $_{-0.004}^{+0.004}$  \\
$-0.25$ & $~0.004$ & $\pm 0.010$ & $_{-0.005}^{+0.006}$  \\
\hline
\end{tabular}
\caption{The values of $\rm \langle\cos\phi^{HCM}\rangle$ and 
         $\rm \langle\cos2\phi^{HCM}\rangle$, calculated using the energy-flow method 
	 as in Eq.~(\ref{eq:method}), as a function of hadron pseudorapidity, 
	 $\rm \eta^{HCM}$. They were obtained in the HCM frame 
	 for the kinematic region \mbox{100 $< Q^2 <$ 8000\,GeV$^2$}, $0.01<x<0.1$ 
	 and $0.2<y<0.8$ for hadrons with $p_T>0.15\,GeV$ and $\theta > 8^\circ$. 
	 The quantities $\delta_{\rm stat}$ and $\delta_{\rm syst}$ are respectively
         the statistical and systematic uncertainties.
}
\label{tab-asym.cos}
\end{center}
\end{table}

\newpage


\begin{table}[!htb]
\begin{center}
\begin{tabular}{||c|c c c ||}
\hline
$\eta^{\rm HCM}$ &$\langle\sin{\phi^{\rm HCM}}\rangle$
   & $\delta_{\rm stat}$ & $\delta_{\rm syst}$  \\
\hline \hline
$-4.75$ & $-0.007$ & $\pm 0.015$ & $_{-0.017}^{+0.006}$  \\
$-4.25$ & $-0.018$ & $\pm 0.010$ & $_{-0.014}^{+0.005}$  \\
$-3.75$ & $-0.023$ & $\pm 0.008$ & $_{-0.002}^{+0.004}$  \\
$-3.25$ & $-0.016$ & $\pm 0.007$ & $_{-0.002}^{+0.004}$  \\
$-2.75$ & $-0.001$ & $\pm 0.007$ & $_{-0.004}^{+0.002}$  \\
$-2.25$ & $-0.012$ & $\pm 0.007$ & $_{-0.001}^{+0.002}$  \\
$-1.75$ & $~0.009$ & $\pm 0.008$ & $_{-0.005}^{+0.001}$  \\
$-1.25$ & $-0.003$ & $\pm 0.008$ & $_{-0.001}^{+0.006}$  \\
$-0.75$ & $-0.018$ & $\pm 0.009$ & $_{-0.004}^{+0.006}$  \\
$-0.25$ & $-0.025$ & $\pm 0.010$ & $_{-0.006}^{+0.003}$  \\
\hline
\multicolumn{4}{c}{~~~}\\
\hline
$\eta^{\rm HCM}$ &$\langle\sin{2\phi^{\rm HCM}}\rangle$
   & $\delta_{\rm stat}$ & $\delta_{\rm syst}$  \\
\hline \hline
$-4.75$ & $~0.012$ & $\pm 0.015$ & $_{-0.002}^{+0.010}$  \\
$-4.25$ & $-0.004$ & $\pm 0.010$ & $_{-0.007}^{+0.001}$  \\
$-3.75$ & $~0.006$ & $\pm 0.008$ & $_{-0.007}^{+0.001}$  \\
$-3.25$ & $-0.009$ & $\pm 0.007$ & $_{-0.002}^{+0.005}$  \\
$-2.75$ & $-0.010$ & $\pm 0.007$ & $_{-0.006}^{+0.004}$  \\
$-2.25$ & $~0.005$ & $\pm 0.007$ & $_{-0.003}^{+0.002}$  \\
$-1.75$ & $-0.009$ & $\pm 0.008$ & $_{-0.002}^{+0.014}$  \\
$-1.25$ & $-0.001$ & $\pm 0.008$ & $_{-0.007}^{+0.006}$  \\
$-0.75$ & $~0.007$ & $\pm 0.010$ & $_{-0.005}^{+0.003}$  \\
$-0.25$ & $~0.010$ & $\pm 0.010$ & $_{-0.002}^{+0.005}$  \\
\hline
\end{tabular}
\caption{The values of $\rm \langle\sin\phi^{HCM}\rangle$ and 
         $\rm \langle\sin2\phi^{HCM}\rangle$, calculated using the energy-flow method 
	 as in Eq.~(\ref{eq:method}), as a function of hadron pseudorapidity, 
	 $\rm \eta^{HCM}$. They were obtained in the HCM frame 
	 for the kinematic region \mbox{100 $< Q^2 <$ 8000\,GeV$^2$}, $0.01<x<0.1$ 
	 and $0.2<y<0.8$ for hadrons with $p_T>0.15\,GeV$ and $\theta > 8^\circ$. 
	 The quantities $\delta_{\rm stat}$ and $\delta_{\rm syst}$ are respectively
         the statistical and systematic uncertainties.
}
\label{tab-asym.sin}
\end{center}
\end{table}

\newpage

\renewcommand{\arraystretch}{1.1}

\begin{table}[!ht]
\begin{center}
\begin{tabular}{||c|c c c ||}
\multicolumn{4}{l}{$\rm -5<\eta^{HCM}\leq -2.5$} \\
\hline
$E_{T, {\rm min}}^{\rm HCM}$\,(GeV) &$\langle\cos{\phi^{\rm HCM}}\rangle$
   & $\delta_{\rm stat}$ & $\delta_{\rm syst}$  \\
\hline \hline
$0.0$ & $-0.064$ & $\pm 0.004$ & $_{-0.005}^{+0.004}$  \\
$0.5$ & $-0.074$ & $\pm 0.005$ & $_{-0.003}^{+0.006}$  \\
$1.0$ & $-0.090$ & $\pm 0.007$ & $_{-0.003}^{+0.011}$  \\
$1.5$ & $-0.099$ & $\pm 0.009$ & $_{-0.007}^{+0.013}$  \\
$2.0$ & $-0.108$ & $\pm 0.011$ & $_{-0.004}^{+0.011}$  \\
$2.5$ & $-0.123$ & $\pm 0.013$ & $_{-0.004}^{+0.015}$  \\
$3.0$ & $-0.128$ & $\pm 0.016$ & $_{-0.007}^{+0.021}$  \\
$3.5$ & $-0.131$ & $\pm 0.019$ & $_{-0.010}^{+0.020}$  \\
$4.0$ & $-0.126$ & $\pm 0.023$ & $_{-0.008}^{+0.019}$  \\
\hline
\multicolumn{4}{c}{~~~}\\
\multicolumn{4}{l}{$\rm -2.5<\eta^{HCM}\leq -1$} \\
\hline
$E_{T, {\rm min}}^{\rm HCM}$\,(GeV) &$\langle\cos{\phi^{\rm HCM}}\rangle$
   & $\delta_{\rm stat}$ & $\delta_{\rm syst}$  \\
\hline \hline
$0.0$ & $~0.000$ & $\pm 0.004$ & $_{-0.012}^{+0.002}$  \\
$0.5$ & $~0.018$ & $\pm 0.006$ & $_{-0.007}^{+0.003}$  \\
$1.0$ & $~0.020$ & $\pm 0.008$ & $_{-0.008}^{+0.004}$  \\
$1.5$ & $~0.017$ & $\pm 0.010$ & $_{-0.010}^{+0.004}$ \\
$2.0$ & $~0.017$ & $\pm 0.011$ & $_{-0.012}^{+0.005}$  \\
$2.5$ & $~0.020$ & $\pm 0.013$ & $_{-0.014}^{+0.005}$ \\
$3.0$ & $~0.018$ & $\pm 0.015$ & $_{-0.015}^{+0.005}$  \\
\hline
\multicolumn{4}{c}{~~~}\\
\multicolumn{4}{l}{$\rm -1<\eta^{HCM}\leq 0$} \\
\hline
$E_{T, {\rm min}}^{\rm HCM}$\,(GeV) &$\langle\cos{\phi^{\rm HCM}}\rangle$
   & $\delta_{\rm stat}$ & $\delta_{\rm syst}$  \\
\hline \hline
$0.0$ & $~0.024$ & $\pm 0.007$ & $_{-0.004}^{+0.004}$  \\
$0.5$ & $~0.037$ & $\pm 0.010$ & $_{-0.003}^{+0.004}$  \\
$1.0$ & $~0.042$ & $\pm 0.013$ & $_{-0.005}^{+0.005}$  \\
$1.5$ & $~0.046$ & $\pm 0.016$ & $_{-0.009}^{+0.007}$  \\
$2.0$ & $~0.044$ & $\pm 0.019$ & $_{-0.009}^{+0.009}$  \\
\hline
\end{tabular}
\vspace*{0.5cm}
\caption{The values of $\rm \langle\cos\phi^{HCM}\rangle$ calculated using the 
         energy-flow method 
         as in Eq.~(\ref{eq:method}), as a function of hadron minimum transverse energy,  
	 $E_{T,min}^{HCM}$. They were obtained in the HCM frame for the pseudorapidity 
	 intervals $-5<\eta^{HCM}\leq -2.5$, $-2.5<\eta^{HCM}\leq -1$ and 
	 $-1<\eta^{HCM}\leq 0$ in the kinematic region \mbox{100 $< Q^2 <$ 8000\,GeV$^2$}, 
	 $0.01<x<0.1$ and $0.2<y<0.8$ for hadrons with $p_T>0.15\,GeV$ and 
	 $\theta > 8^\circ$. 
	 The quantities $\delta_{\rm stat}$ and $\delta_{\rm syst}$ are respectively
         the statistical and systematic uncertainties.
         }
\label{tab-ethcm.cos}
\end{center}
\end{table}

\newpage



\begin{table}[!ht]
\begin{center}
\begin{tabular}{||c|c c c ||}
\multicolumn{4}{l}{$\rm -5<\eta^{HCM}\leq -2.5$} \\
\hline
$E_{T, {\rm min}}^{\rm HCM}$\,(GeV) &$\langle\cos{2\phi^{\rm HCM}}\rangle$
   & $\delta_{\rm stat}$ & $\delta_{\rm syst}$  \\
\hline \hline
$0.0$ & $-0.002$ & $\pm 0.004$ & $_{-0.003}^{+0.003}$  \\
$0.5$ & $~0.002$ & $\pm 0.005$ & $_{-0.002}^{+0.003}$  \\
$1.0$ & $~0.006$ & $\pm 0.007$ & $_{-0.002}^{+0.002}$  \\
$1.5$ & $~0.009$ & $\pm 0.009$ & $_{-0.010}^{+0.002}$  \\
$2.0$ & $~0.016$ & $\pm 0.011$ & $_{-0.018}^{+0.001}$  \\
$2.5$ & $~0.014$ & $\pm 0.014$ & $_{-0.016}^{+0.003}$  \\
$3.0$ & $~0.014$ & $\pm 0.017$ & $_{-0.028}^{+0.004}$  \\
$3.5$ & $~0.019$ & $\pm 0.021$ & $_{-0.010}^{+0.005}$  \\
$4.0$ & $~0.006$ & $\pm 0.025$ & $_{-0.005}^{+0.012}$  \\
\hline
\multicolumn{4}{c}{~~~}\\
\multicolumn{4}{l}{$\rm -2.5<\eta^{HCM}\leq -1$} \\
\hline
$E_{T, {\rm min}}^{\rm HCM}$\,(GeV) &$\langle\cos{2\phi^{\rm HCM}}\rangle$
   & $\delta_{\rm stat}$ & $\delta_{\rm syst}$  \\
\hline \hline
$0.0$ & $~0.022$ & $\pm 0.004$ & $_{-0.007}^{+0.004}$  \\
$0.5$ & $~0.039$ & $\pm 0.006$ & $_{-0.006}^{+0.005}$  \\
$1.0$ & $~0.055$ & $\pm 0.008$ & $_{-0.009}^{+0.006}$  \\
$1.5$ & $~0.062$ & $\pm 0.009$ & $_{-0.011}^{+0.007}$  \\
$2.0$ & $~0.067$ & $\pm 0.011$ & $_{-0.010}^{+0.008}$  \\
$2.5$ & $~0.070$ & $\pm 0.012$ & $_{-0.010}^{+0.009}$  \\
$3.0$ & $~0.076$ & $\pm 0.014$ & $_{-0.011}^{+0.010}$  \\
\hline
\multicolumn{4}{c}{~~~}\\
\multicolumn{4}{l}{$\rm -1<\eta^{HCM}\leq 0$} \\
\hline
$E_{T, {\rm min}}^{\rm HCM}$\,(GeV) &$\langle\cos{2\phi^{\rm HCM}}\rangle$
   & $\delta_{\rm stat}$ & $\delta_{\rm syst}$  \\
\hline \hline
$0.0$ & $~0.018$ & $\pm 0.007$ & $_{-0.004}^{+0.005}$  \\
$0.5$ & $~0.022$ & $\pm 0.009$ & $_{-0.005}^{+0.005}$  \\
$1.0$ & $~0.026$ & $\pm 0.013$ & $_{-0.008}^{+0.008}$  \\
$1.5$ & $~0.032$ & $\pm 0.016$ & $_{-0.010}^{+0.010}$  \\
$2.0$ & $~0.034$ & $\pm 0.018$ & $_{-0.013}^{+0.012}$  \\
\hline
\end{tabular}
\vspace*{0.5cm}
\caption{The values of $\rm \langle\cos\phi^{HCM}\rangle$ calculated using the 
         energy-flow method 
         as in Eq.~(\ref{eq:method}), as a function of hadron minimum transverse energy,  
	 $E_{T,min}^{HCM}$. They were obtained in the HCM frame for the pseudorapidity 
	 intervals $-5<\eta^{HCM}\leq -2.5$, $-2.5<\eta^{HCM}\leq -1$ and 
	 $-1<\eta^{HCM}\leq 0$ in the kinematic region \mbox{100 $< Q^2 <$ 8000\,GeV$^2$}, 
	 $0.01<x<0.1$ and $0.2<y<0.8$ for hadrons with $p_T>0.15\,GeV$ and 
	 $\theta > 8^\circ$. 
	 The quantities $\delta_{\rm stat}$ and $\delta_{\rm syst}$ are respectively
         the statistical and systematic uncertainties.
         }
\label{tab-ethcm.cos2}
\end{center}
\end{table}


%
%
\end{document}